# Hardware Trojan Threats to Multi-Chiplet Photonic Neural Network Accelerators

Sudeep Pasricha
Department of Electrical and Computer Engineering
Colorado State University
sudeep@colostate.edu

*Special Session Paper*

***Abstract*—Multi-chiplet photonic neural network accelerators (MCPNAs) combine photonic computation, photonic communication, and heterogeneous chiplet integration to enable scalable and energy-efficient AI acceleration. However, their distributed architecture and reliance on third-party chiplets introduce significant hardware security risks. This paper examines Hardware Trojan (HT) threats to MCPNAs across three dimensions: confidentiality, integrity, and availability.**

## I. Introduction

The adoption of 2.5D chiplet integration provides a promising path toward scaling photonic neural network accelerators beyond the limits of monolithic designs. In multi-chiplet photonic neural network accelerator (MCPNA) architectures, such as our work in [1] (see Fig. 1) and that of others [2], [3], photonic compute chiplets communicate with memory chiplets through a silicon-photonic interposer using wavelength-division multiplexing (WDM). The architecture relies on microring resonators (MRs), photonic gateways, wavelength-routing structures, photodetectors, and thermo/electro-optic tuning circuits [4] distributed across chiplets and the interposer.

While this organization enables scalable and energy-efficient AI acceleration, MCPNAs also expose a larger attack surface than monolithic accelerators, across the supply chain. We consider a threat model where Hardware Trojans (HTs) are introduced into untrusted compute, memory, gateway, or interposer chiplets during design, fabrication, or integration. HTs may manipulate microring tuning or thermal control, snoop optical signals, or disrupt inter-chiplet traffic, independently or through collusion. The resulting perturbations are difficult to detect due to the inherent sensitivity of photonic devices to wavelength, temperature, and process variations. The security objectives are therefore to protect model confidentiality, integrity, and system availability.

## II. Confidentiality Attacks

A first class of attacks targets confidentiality. The MCPNA employs shared optical waveguides carrying multiple WDM channels between chiplets. In SOTERIA [5], [6], we showed that HTs embedded in MR tuning circuits can maliciously retune detector MRs to partially resonate with neighboring wavelengths, enabling covert snooping of optical channels with little impact on legitimate communication. In an MCPNA, such attacks can reveal weight values fetched from memory chiplets, intermediate activations exchanged between compute chiplets, and proprietary model parameters. As future systems increasingly integrate third-party chiplets, such leakage poses a significant threat to model confidentiality and intellectual property protection.

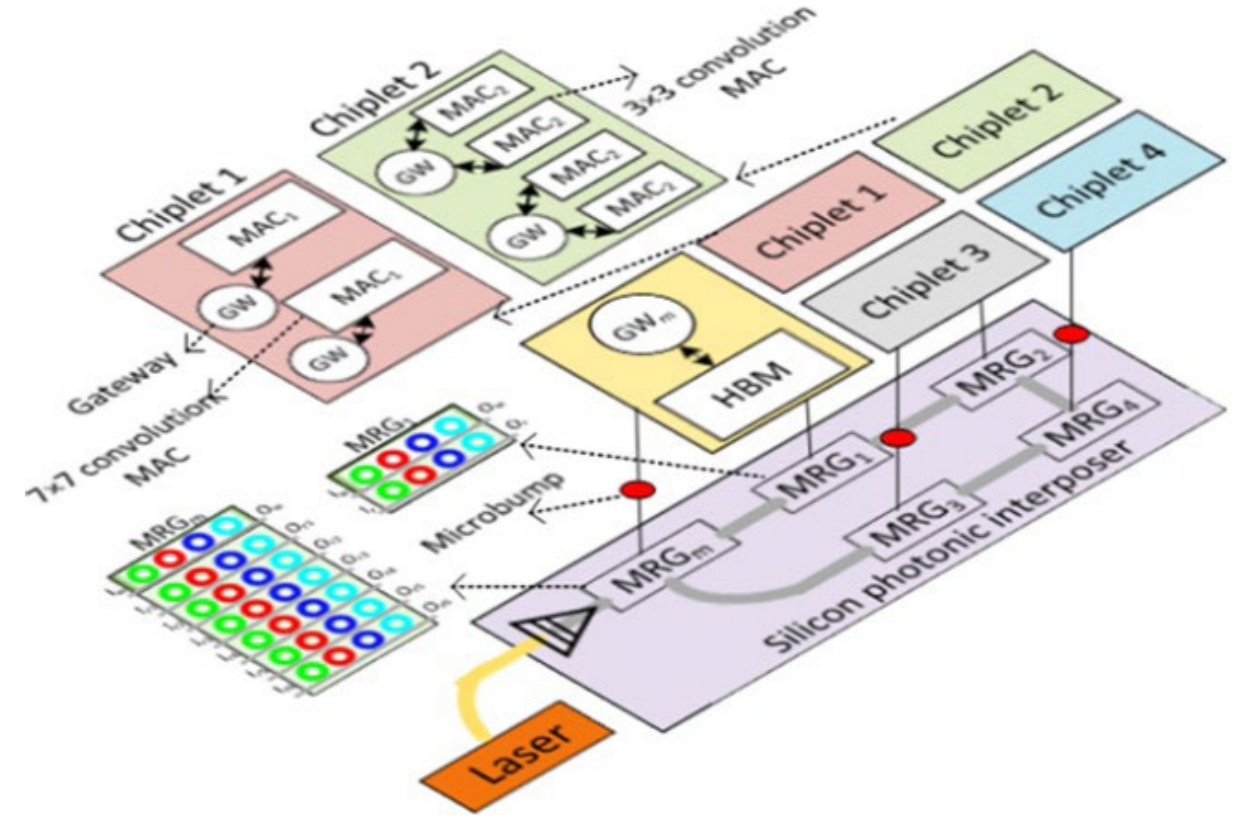

Fig. 1: Multi-chiplet photonic neural network accelerator (MCPNA) architecture [1]. Such architectures are increasingly susceptible to Hardware Trojan (HT) attacks that compromise confidentiality, integrity, and availability.

SOTERIA further demonstrated that photonic process variations can be exploited as a security asset rather than merely a reliability challenge. The framework derives unclonable authentication and encryption keys from process-variation signatures of photonic devices, protecting communications from snooping attacks. In future MCPNAs, these ideas could be extended to chiplet authentication, trusted onboarding, and secure photonic communication protocols. Process-variation-based photonic fingerprints offer a particularly attractive mechanism for establishing trust in externally sourced chiplets.

## III. Integrity Attacks

A second attack class targets the integrity of photonic neural computations. The MCPNA in [1] extends the CrossLight architecture [7] by distributing photonic MAC units across multiple compute chiplets, with MRs encoding neural-network weights and activations. As shown in SafeLight [8], HTs inserted into MR actuation circuitry can force devices into off-resonance states, corrupting weight or activation encoding and causing incorrect computations. As neural-network layers are distributed across chiplets, errors originating in one chiplet can propagate through subsequent computations and inter-chiplet communications, impacting the entire inference pipeline.

SafeLight also identified thermal hotspot attacks as an especially potent HT payload. By manipulating thermo-optic tuning circuits, malicious circuitry can generate localized heating that shifts resonance wavelengths for multiple MRs. In MCPNAs, where dense photonic MAC arrays share a common interposer substrate, thermal coupling can extend the impact of an attack far beyond the compromised device. A Trojan in one compute chiplet may therefore influence neighboring MRs or even adjacent chiplets. Ultimately, it was shown that even when only a small fraction of MRs are compromised, significant accuracy degradation can occur. Further, the thermal hotspot attacks often cause larger accuracy degradation than actuation attacks because they simultaneously corrupt clusters of parameters rather than isolated weights.

Mitigating integrity attacks requires both hardware and algorithmic resilience. SafeLight demonstrated that L2 regularization and noise-aware training significantly improve tolerance to HT-induced perturbations. In future MCPNAs, these techniques could be combined with runtime thermal monitoring, resonance tracking, hardware redundancy, and chiplet-level fault containment. Such cross-layer approaches would allow neural networks to tolerate device-level attacks while enabling runtime detection of abnormal photonic behavior.

## IV. Availability Attacks

A third class of attacks targets system availability. MCPNAs depend on high-bandwidth inter-chiplet communication through photonic gateways and interposer networks, making these resources attractive targets. SCRIPT [9] showed how malicious chiplets can coordinate distributed denial-of-service (DDoS) attacks by concentrating traffic on critical communication resources. Although originally developed for electronic interposers, the threat model applies naturally to photonic systems. A compromised chiplet could inject excessive traffic, monopolize wavelength channels, overwhelm gateway interfaces, or coordinate with other malicious chiplets to create congestion on critical optical paths.

MCPNAs are particularly vulnerable because communication between memory and compute chiplets often follows predictable routes through dedicated gateways. An attacker exploiting this predictability can create congestion hotspots that increase latency and reduce accelerator utilization. Reconfigurable photonic interposers may introduce additional vulnerabilities if HTs manipulate gateway or wavelength allocation decisions to reduce effective bandwidth.

SCRIPT points toward trust-aware routing as a promising defense. By incorporating chiplet trust levels into routing decisions and introducing path obfuscation, randomized route selection, and load balancing, the framework reduces an adversary's ability to exploit critical communication paths. In photonic chiplet accelerators, these principles can be extended to wavelength assignment, gateway selection, and photonic path management through techniques such as dynamic wavelength hopping and adaptive interposer reconfiguration.

## V. Conclusions

Multi-chiplet photonic neural network accelerators (MCPNAs) offer a scalable and energy-efficient platform for future AI workloads, but their distributed architecture introduces new Hardware Trojan (HT) attack surfaces. This paper highlighted how HTs can compromise confidentiality through photonic interconnect snooping, integrity through microring actuation and thermal attacks, and availability through coordinated denial-of-service attacks on inter-chiplet communication resources. We discussed emerging countermeasures including photonic authentication, resilient model training, runtime monitoring, and trust-aware routing. These observations underscore the need for holistic cross-layer security frameworks to ensure trustworthy operation of next-generation photonic chiplet-based AI systems. As future photonic AI platforms scale to support workloads such as LLMs [10], graph learning [11], and vision-based generative AI [12], effective protection will require cross-layer defenses spanning trusted chiplet onboarding, secure photonic communication, resilient model training, runtime anomaly detection, thermal management, and trust-aware network orchestration.

## VI. Acknowledgements

This research was supported in part by National Science Foundation grant CCF-2450615.